\documentclass[12pt,a4paper]{article}
 \usepackage{epsf}
\begin{document}
\title{The rare top decay $t\rightarrow c g$ in
topcolor assisted technicolor models}
\author{Chongxing Yue$^{(a,b)}$, Gongru Lu$^{(a,b)}$,
 Guoli Liu$^{b}$, Qingjun Xu$^{b}$
 \\ {\small a: CCAST (World
 Laboratory) P.O. BOX 8730. B.J. 100080 P.R. China} \\
 {\small b:College of Physics and Information Engineering,}\\
 \small{Henan Normal University,
 Xinxiang  453002. P.R.China}
\thanks{This work is supported by the National Natural Science
 Foundation of China(I9905004), the Excellent Youth Foundation of
 Henan Scientific Committee(9911); and Foundation of Henan
 Educational Committee.}
\thanks{E-mail:cxyue@public.xxptt.ha.cn} }
\date{\today}
\maketitle
\begin{abstract}
\hspace{5mm} In the framework of topcolor-assisted
technicolor(TC2) models, we calculate the contributions
of flavor
changing scalar couplings involved the neutral top-pion
$\pi_{t}^{0}$ and top-Higgs $h_{t}^{0}$ to the branching ratio
$B_{r}(t\longrightarrow cg)$. We find that the value of
$B_{r}(t\longrightarrow cg)$ can reach $ 1\times10^{-3}$ with
reasonable values of the parameters in TC2 models, which may be
testable in the future experiments.
\end {abstract}
\vspace{1.0cm}

\newpage

   It is widely believed that the top quark, with a mass of the
order of the electroweak scale, will be a sensitive probe into
physics beyond the standard(SM). Indeed, the properties of the top
quark could reveal information of flavor physics, electroweak
symmetry breaking(EWSB) as well as new physics beyond the SM
\cite{b1}. One of these consequences is that the rare top decays
can be used to detect new physics. A lot of the theoretical
activity involving the rare top decays have been given within some
specific models beyond the SM \cite{b2}.

    At the tree-level there are no flavor changing neutral
currents processes in the SM, and at one-loop they are induced by
charged current interactions, which are GIM-suppressed. In
particular, the rates for the flavor changing rare decays of the
top quark $ t\longrightarrow cv(v=g,\gamma, Z)$ are very small.
With the current experimental value of the top quark mass, the
branching ratios are $B_{r}(t\longrightarrow cg)\sim 4\times
10^{-13}$, $B_{r}(t\longrightarrow c\gamma)\sim 5\times 10^{-13}$
and $B_{r}(t\longrightarrow cZ)\sim 1\times 10^{-13}$ in the
SM \cite{b3}, which are far below the feasible experimental
possibilities at the future colliders(LHC or LC)\cite{b4}. The
values of the branching ratios $B_{r}(t\longrightarrow
cvv)(v=W,g,\gamma, or Z) $ predicted by the SM are more small.
These have been studied in Ref.[5]. It is widely believed that, in
many simple SM extensions, the branching ratios
$B_{r}(t\longrightarrow cv)$ can be enhanced by several orders of
magnitude. Detection of the rare top decays at visible levels by
any of the future colliders would be instant evidence of new
physics. This fact has lead to a lot of studies involving the rare
top decays within some specific models beyond the SM. For
instance, studies of the rare top decays $B_{r}(t\longrightarrow
cv)$ in the multi Higgs doublets models(MHDM) \cite{b3,b6}, in
supersymmetry with R-Parity conservation\cite{b7} and with
R-Parity violation \cite{b8}, and in technicolor models \cite{b9}.

   There are many ways in which the extensions of the SM  lead to
flavor changing scalar(FCS) couplings at the tree-level. FCS
couplings can normally enhance the values of branching ratios
$B_{r}(t\longrightarrow cv)$. This has been shown that the FCS
couplings predicted by the general 2HDM typeIII can significantly
enhance the branching ratios of the rare top decays
$t\longrightarrow cv$ \cite{b10}. Recently, Ref.[11] has
calculated the branching ratio for $t\longrightarrow c\gamma$ in
the framework of the most general CP-conserving 2HDM typeIII. They
have shown that, with reasonable values for the parameter
$tan\beta$, the branching ratio $B_{r}(t\longrightarrow cv)$ could
be within the observable threshold of near future experiments. The
aim of this paper is to point out that the FCS couplings predicted
by topcolor assisted technicolor(TC2) models \cite{b12} also can
significantly enhance the values of the branching ratios
$B_{r}(t\longrightarrow cv)$, which may approach the detectability
threshold of near future experiments.

  An important issue in high-energy physics is to understand the
mechanism of the mass generation. There may be a common origin for
EWSB and top quark mass generation. Much theoretical work has been
carried out in connection to the top quark and EWSB. TC2 models
\cite{b12} and the top see-saw models \cite{b13} are two of such
examples. Such type of models generally predict a number of
scalars with large Yukawa coupling to the third generation
fermions. For example, TC2 models predict the existence of scalars
including the technipions $(\pi^{0}, \pi^{\pm}; \pi_{a}^{0} ,
\pi_{a}^{\pm})$ in the technicolor sector and the top-pions
$(\pi_{t}^{0},\pi_{t}^{\pm})$ and top-Higgs $h_{t}^{0}$ in the
topcolor sector. Ref.[9] has considered the contributions of these
new particles to the rare top decays $t \longrightarrow cv$ ,
which gave $B_{r}(t \longrightarrow c g) \sim 10^{-6}$,
$B_{r}(t\longrightarrow c \gamma) \sim 10^{-7}$ and $B_{r}(t
\longrightarrow c Z) \sim 10^{-8}$. However, Ref.[9] only
considered the contributions of charged scalars ($\pi^{\pm}$,
$\pi^{\pm}_{a}$ and $\pi_{t}^{\pm}$) via the couplings
$s^{\pm}u_{i}u_{j}$ and did not consider the contributions of the
neutral top-pion $\pi_{t}^{0}$ and top-Higgs $h_{t}^{0}$ via the
FCS coupling $s \overline{t}c$, in which, s is behalf of
$\pi_{t}^{0}$ or $h_{t}^{0}$. In fact, it is an important feature
of TC2 models that the neutral scalars can induce the tree-level
FCS couplings. It has been shown that the FCS couplings can give
distinct new flavor mixing phenomena which may be tested at both
low and high energy experiments \cite{b14,b15}. Thus, in this
paper, we will calculate the contributions of the FCS coupling
$s\overline{t}c$ to the rare top decays $t\longrightarrow c v$ in
the framework of TC2 models and see whether $t\longrightarrow c v$
can be used to test TC2 models.

   Ref.[9] has shown that the branching ratios  $B_{r}
   (t\longrightarrow c
v)$ can be enhanced several orders of magnitude by charged
scalars. The rare top decays $t\longrightarrow c v(v=\gamma, Z)$
are far below the observable threshold of near future experiments.
However, the branching ratio $B_{r}(t\longrightarrow c g)$ can
reach to $10^{-5}$ for the favorable parameter values of TC2
models, which is just below the expected experimental sensibility.
Thus $t\longrightarrow c g$ is the most promising one among the
rare top decays into gauge bosons $t\longrightarrow c v(v=g,
\gamma, Z)$. Certainly, we must separate the signals from the
large backgrounds before observation of the rare decay
$t\longrightarrow c g$ at the future LHC experiments. However, a
future high-energy linear $e^{+}e^{-}$ collider (LC) could, in
principle, be the ideal place to study $t\longrightarrow c g$, as
top quark events can be clearly separated by tagging the isolated
lepton from the top semileptonic decay. Perhaps, with high
integrated luminosity, the largest decay channel $t\longrightarrow
c g$ can be detected at LC. Thus, our attention will mainly focus
on the rare top decay $t\longrightarrow c g$. Our results show
that the branching ratio $B_{r}(t\longrightarrow c g)$ may
approach the observable threshold of near future experiments. For
$\epsilon=0.08$, $m_{\pi_{t}}=200GeV$, the branching ratio
$B_{r}(t\longrightarrow c g)$ contributed by the neutral top-pion
$\pi_{t}^{0}$ can reach $6\times 10^{-4}$ and  for
$\epsilon=0.08$, $m_{h_{t}}=200GeV$, we have
$B_{r}(t\longrightarrow c g)\approx 1\times 10^{-3}$ arised from
the top-Higgs $h_{t}^{0}$, which may be the order of sensitivity
of the future experiments LHC or LC.

    In TC2 models, the TC interactions play a main role in
breaking the electroweak gauge symmetry. The ETC interactions give
rise to the masses of the ordinary fermions including a very small
portion of the top quark mass, namely $\epsilon m_{t}$ with a
model dependent parameter $\epsilon \ll 1$. The topcolor
interactions also make small contributions to the EWSB, and give
rise to the main part of the top quark mass, $(1-\epsilon) m_{t}$,
similar to the constituent masses of the light quarks in QCD. So
that the heaviness of the top-quark emerges naturally in TC2
models. This kind of models predict three top-pions with large
Yukawa couplings to the third generation. This induces the new FCS
couplings. The relevant couplings including the t-c transition for
the neutral top-pion $\pi_{t}^{0}$ can be written as
\cite{b12,b14}:
\begin{equation}
\frac{m_{t}}{\sqrt{2}F_{t}}\frac{\sqrt{\nu_{w}^{2}-F_{t}^{2}}}
{\nu_{w}}
[K_{UR}^{tt}K_{UL}^{tt*}\bar{t_{L}}t_{R}\pi_{t}^{0}+K_{UR}^{tc}
K_{UL}^{tt*}\bar{t_{L}}c_{R}\pi_{t}^{0}+h.c.],
\end{equation}
where the factor $\sqrt{\nu_{w}^{2}-F_{t}^{2}}/\nu_{w}$
reflects
the effects of the mixing between the neutral top-pion
$\pi_{t}^{0}$ and the would be Goldstone boson with
$\nu_{w}=v/\sqrt{2}=174GeV$ and $F_{t}=50GeV$ which is the
top-pion decay constant. $k_{UL}^{ij}$ is the matrix element of
unitary matrix $k_{UR}^{ij}$ is the matrix element of the
right-handed relation matrix $k_{UR}$. Ref.[14] has shown that
their values can be taken as:
\begin{equation}
K_{UL}^{tt}=1, \hspace{5mm}  K_{UR}^{tt}=1-\epsilon, \hspace{5mm}
K_{UR}^{tc}\leq \sqrt{2\epsilon-\epsilon^{2}}.
\end{equation}
In the following calculation, we will take $K_{UR}^{tc}=
\sqrt{2\epsilon-\epsilon^{2}}$ and take $\epsilon$ as a free
parameter.

  The relevant Feynman diagrams for the contributions of the
neutral top-pion $\pi_{t}^{0}$ to the rare top decay
$t\longrightarrow cg$ via the FCS coupling are shown in Fig.1.
Using Eq.[1], we can give the decay width $\Gamma(t\longrightarrow
cg)$:

\begin{eqnarray}
\nonumber \Gamma(t\longrightarrow
cg)&=&\frac{\alpha_{s}m_{t}^{5}}{512\pi^{4}
F_{t}^{4}}\frac{v_{w}^{2}-F_{t}^{2}}{v_{w}^{2}}(K^{tc})^{2}
[m_{t}^2A_{1}(C_{12}-2C_{21}+3C_{23}-C_{22})+2A_{1}^{2}\\
\nonumber &&+2m_{t}^{4}(-C_{12}C_{21}-C_{12}C_{22}
+2C_{12}C_{23}+C_{21}^{2}+C_{21}C_{22}-3C_{21}C_{23}\\
&&-C_{22}C_{23}+2C_{23}^{2})]
\end{eqnarray}
where
\begin{equation}
A_{1}=m_{t}^{2}(C_{11}-C_{12})
-2C_{24}+B_{0}+m_{\pi}^{2}C_{0}+\frac{m_{c}}{m_{t}-m_{c}}B_{1}
-B_{0}^{*}-\frac{m_{t}}{m_{t}-m_{c}}B_{1}^{*},
\end{equation}
with
\begin{equation}
B_{0}=B_{0}(p_{g},m_{t},m_{t},m_{\mu}),\hspace{8mm}
B_{0}^{*}=B_{0}(-p_{c},m_{s},m_{t},m_{\mu}),
\end{equation}
\begin{equation}
B_{1}=B_{1}(-p_{c},m_{s},m_{t},m_{\mu}), \hspace{8mm}
B_{1}^{*}=B_{1}(-p_{t},m_{\pi},m_{t}),
\end{equation}
\begin{eqnarray}
 C_{0}&=&C_{0}(-p_{t},m_{s},m_{t},m_{\mu}),   \hspace{5mm}
C_{24}=C_{24}(-p_{t},p_{g},m_{s},m_{t},m_{t},m_{\mu}),
 \hspace{3mm}\\
&& C_{ij}=C_{ij}(-p_{t},m_{s},m_{t},m_{\mu}),\hspace{5mm}
i,j=1,2,3.
\end{eqnarray}
$B_{0}$, $B_{0}^{*}$, $B_{1}$, $B_{1}^{*}$, $C_{0}$ $C_{24}$ and
$C_{ij}$ are standard feymman integrals, in which variable $p_{c}$
is the monument of charm quarks,
 $p_{t}$ is the monument of top quarks,
 $p_{g}$ is the monument of the gluons and $m_{\mu}$ is the scale
  of the TC2 models.
  Since the $\overline{c}_{L}t_{R}$
coupling is very small \cite{b14}, we have assumed
$K_{UR}^{tc}\approx K^{tc}=
\sqrt{|K_{UL}^{tc}|^{2}+|K_{UR}^{tc}|^{2}}$ in the above
equations.

  Ref.[12] has estimated the mass of the top-pion in the fermion
loop approximation and given $180GeV\leq m_{\pi_{t}}\leq 250GeV$
for $m_{t}=180GeV$ and $0.03 \leq \epsilon \leq 0.1$. Since the
negative top-pion corrections to the $Z\longrightarrow
b\overline{b}$ branching ratio $R_{b}$ becomes smaller when the
top-pion is heavier, the LEP/SLD data of $R_{b}$ give rise to
certain lower bound on the top-pion mass \cite{b16}. It was shown
that the top-pion mass should not be lighter than the order of
1TeV to make the TC2 models consistent with the LEP/SLD data.
However, we restudied this problem in Ref.[17]. Our results show
that the top-pion mass $m_{\pi_{t}}$ is allowed to be in the
region of a few hundred GeV depending on the models. Thus, the
top-pion mass depends on the value of the parameters in TC2
models. As estimation the contributions of the neutral top-pion
$\pi_{t}^{0}$ to the rare top decay $t\longrightarrow cg$, we take
the mass of $\pi_{t}^{0}$ to vary in the range of $200GeV-400GeV$
in this paper.

   The branching ratio $B_{r}(t\longrightarrow cg)$ contributed by
the neutral top-pion $\pi_{t}^{0}$ is plotted in Fig.2 as a
function of the top-pion mass $m_{\pi_{t}}$ for three values of
the parameter $\epsilon$. In Fig.2 we have assumed that the total
width is dominated by the decay channel $t\longrightarrow Wb$ and
taken $\Gamma(t\longrightarrow Wb)=1.56GeV$\cite{b1}. The integral
was performed numerically  for the parameter values
$m_{t}=175GeV$, $m_{c}=1.2GeV$ and $\alpha_{s}=0.118$ \cite{b18}.
From Fig.2, we can see that the FCS coupling
$\pi_{t}^{0}\overline{t}c$  indeed could give significantly
contributions to the rare top decay $t\longrightarrow cg$. The
value of $B_{r}(t\longrightarrow cg)$ increases with $\epsilon$
increasing and $m_{\pi_{t}}$ decreasing. For $m_{\pi_{t}}=300GeV$,
the branching ratio  $B_{r}(t\longrightarrow cg)$ varies between
$0.60\times 10^{-4}$ and $3.45\times 10^{-4}$ for the parameter
$\epsilon$ in the range of 0.01-0.08. For $m_{\pi_{t}}=200GeV$ and
$\epsilon=0.08$, the value of $B_{r}(t\longrightarrow cg)$ can
reach $6.1\times 10^{-4}$.

   One possible set of anomalous interactions for top quark is
given  by the flavor-changing chromo-magnetic
operators\cite{b19,b20,b21}:
\begin{equation}
\frac{K_{f}}{\Lambda}g_{s} \overline{f}\sigma^{\mu\nu}
\frac{\lambda^{a}}{2}tG_{\mu\nu}^{a}+h.c.
\end{equation}
where $\Lambda$ is the new physics scale, f=u or c, the $K_{f}$
define the strength of the $ \overline{t}ug$ or $ \overline{t}cg$
anomalous couplings, and $G_{\mu\nu}^{a}$ is the gauge field
tensor of the gloun. Ref.[19] has studied the discovery limits of
$K_{c}/\Lambda$ and $K_{u}/\Lambda$ at Tevatron and LHC using
direct production of an s-channel top quark. The minimum values of
$K_{c}/\Lambda$ and $K_{u}/\Lambda$ observable at Tevatron and LHC
are also calculated in Ref.[20] using single top quark production.
Their results are summarized in Ref.[1] by M.Beneke et.al.
Ref.[19] has shown that the strength of the anomalous coupling
$\overline{t}cg$ may be detected down to $K_{c}=0.03$ at the
Tevatron with $30fb^{-1}$ of data at 2TeV and $K_{c}=0.0084$ at
the LHC with $10fb^{-1}$ of data at 14TeV, and the results of
Ref.[20] are that the anomalous coupling $ \overline{t}cg$ may be
detected down to $K_{c}=0.046$ at the Tevatron with $30fb^{-1}$ of
data at 2TeV and $K_{c}=0.013$ at the LHC with $10fb^{-1}$ of data
at 14TeV.

  If we assume that the anomalous coupling $\overline{t}cg$ comes
from the neutral top-pion $\pi_{t}^{0}$, the contributions of
$\pi_{t}^{0}$ to $Br(t\longrightarrow cg)$ can be transposed to
that of the strength $K_{c}$. According to our calculation, we
have that, for $m_{\pi_{t}}=300GeV$, the value of $K_{c}$
increases from 0.0081 to 0.019 as the parameter $\epsilon$
increases from 0.01 to 0.08. For $m_{\pi_{t}}=200GeV$ and
$\epsilon=0.08$, the value of $K_{c}$ contributed by the neutral
top-pion $\pi_{t}^{0}$ can reach 0.026. Thus, the values of the
branching ratio $Br(t\longrightarrow cg)$ are in the range of the
sensitivity of the future LHC experiments for the favorable
parameter values of TC2 models. The effects of the FCS coupling
$\pi_{t}^{0} \overline{t}c $ predicted by TC2 models on the rare
decay $t\longrightarrow cg$ can not be probed at the Tevatron
experiments, but may be probed at the future LHC experiments.

   TC2 models also predict the existence of the neutral
CP-even state, called top Higgs $h_{t}^{0}$. The main difference
between the neutral top-pion $\pi_{t}^{0}$ and top-Higgs
$h_{t}^{0}$ is that $h_{t}^{0}$ can couple to gauge boson pairs WW
and ZZ at tree level, which is similarly to that of the SM Higgs
$H^{0}$(The couplings $h_{t}^{0}WW$ and $h_{t}^{0}ZZ$ are
suppressed by the factor $F_{t}/\nu_{w}$ with respect to that of
$H^{0}$\cite{b14}.). Thus, the contributions of $h_{t}^{0}$ to the
rare decay $t\longrightarrow cg$ are similar to that of
$\pi_{t}^{0}$. The relevant Feynman diagrams are also shown in
Fig.1. Our results are plotted in Fig.3. To compare our results
with that of Refs[19,20] and see whether the effects of
$h_{t}^{0}$ on the rare decay $t\longrightarrow cg$ can be
detected, we plot the anomalous coupling strength $K_{c}$ as a
function of the top-Higgs mass $m_{h_{t}}$ for the three values of
parameter $\epsilon$ in Fig.3. The other parameter values are the
same as those of Fig.2. From Fig.3, we can see that the
contributions of the top-Higgs $h_{t}^{0}$ to the rare top decay
$t\longrightarrow cg$ also decrease with increasing $m_{h_{t}}$,
which is similar to that of
 the neutral top-pion $\pi_{t}^{0}$. However, the contributions
of $h_{t}^{0}$ to the branching ratio $Br(t\longrightarrow cg)$ is
larger than that of $\pi_{t}^{0}$. For $m_{h_{t}}=300GeV$, the
strength $K_{c}$ varies between 0.011 and 0.026 for the parameter
$\epsilon$ in the range of 0.01-0.08. For $m_{h_{t}}=200GeV$ and
$\epsilon=0.08$, the value of the strength $K_{c}$ can reach
0.033.

   If the effects of the new particles such as $\pi_{t}^{0}$ and
$h_{t}^{0}$ on the rare decay $t\longrightarrow cg$ could be
measured in the future colliders, then one should  consider to
search directly for these new particles. In fact, these has been
done by many authors. For example, G. Burdman \cite{b14} has
studied the observability of the neutral top-pion and top-Higgs at
Tevatron or LHC, and H.-J. He and C. P. Yuan \cite{b14} have
studied the observability of the charged top-pion in the future
colliders. We have shown that the neutral top-pion  may be
detected via the processes $e^{+}e^{-}\longrightarrow
\overline{t}c$, $e^{+}e^{-}\longrightarrow
\overline{t}c\overline{\nu_{e}}\nu_{e}$, and
$e^{+}e^{-}\longrightarrow \gamma \pi_{t}^{0}\longrightarrow
\gamma \overline{t}c$ in the future LC experiments \cite{b15}.
Certainly, the neutral scalars $h_{t}^{0}$, $\pi_{t}^{0}$ can also
be produced via the process $q \overline{q}\longrightarrow g
\longrightarrow \pi_{t}^{0}(h_{t}^{0}) g$. In the future, we will
discuss the possibility of detecting $\pi_{t}^{0}$, $h_{t}^{0}$
via this process in hadron colliders.

  In conclusion, we have conformed that the FCS couplings
predicted by TC2 models can give significantly contributions to
the rare top decay $t\longrightarrow cg$. The branching ratio
$B_{r}(t\longrightarrow cg)$ can be highly enhanced, which is in
the range of the sensitivity of future experiments. Further, it is
worth to remark that this enhancement in TC2 models also appear in
other rare top decays. For instance, the value of the branching
ratio $B_{r}(t\longrightarrow c\gamma)$ varies between $7.95\times
10^{-7}$ and $4.58\times 10^{-6}$ for $m_{h_{t}}=300GeV$ and the
parameter $\epsilon$ in the range of 0.01-0.08, which approach the
corresponding experimental threshold.
\newline
{\bf Acknowledgments}
\newline
Chongxing Yue would like to thank C. P. Yuan for many
 useful discussions.
 \newpage
 \vskip 2.0cm
 \begin{center}
 {\bf Figure Captions}
 \end{center}
 \begin{description}
 \item[Fig.1:]Feynman diagrams for the contributions of the
 neutral top-pion $\pi_{t}^{0}$ or the top-Higgs $h_{t}^{0}$
  via
 the FCS couplings($\pi^{0}_{t}\overline{t}c$ or $h_{t}^{0}
 \overline{t}c$) to the rare decay $t\longrightarrow cg$.
 \item[Fig.2:]The branching ratio $B_{r}(t\longrightarrow cg)$ as
 a function of the neutral top-pion mass $m_{\pi_{t}}$ for the
 parameter $\epsilon=0.08$(solid line), $0.05$(dotted line) and
 $0.01$(dashed line).
 \item[Fig.3:]The strength $K_{c}$ of the anomalous coupling as a
 function of the neutral top-Higgs mass
 $m_{h_{t}}$ for the parameter $\epsilon=0.08$(solid line),
 $0.05$(dotted line) and $0.01$(dashed line).
 \end{description}
 
\newpage
\begin{figure}[pt]
\begin{center}
\begin{picture}(250,300)(0,0)
\put(-50,0){\epsfxsize120mm\epsfbox{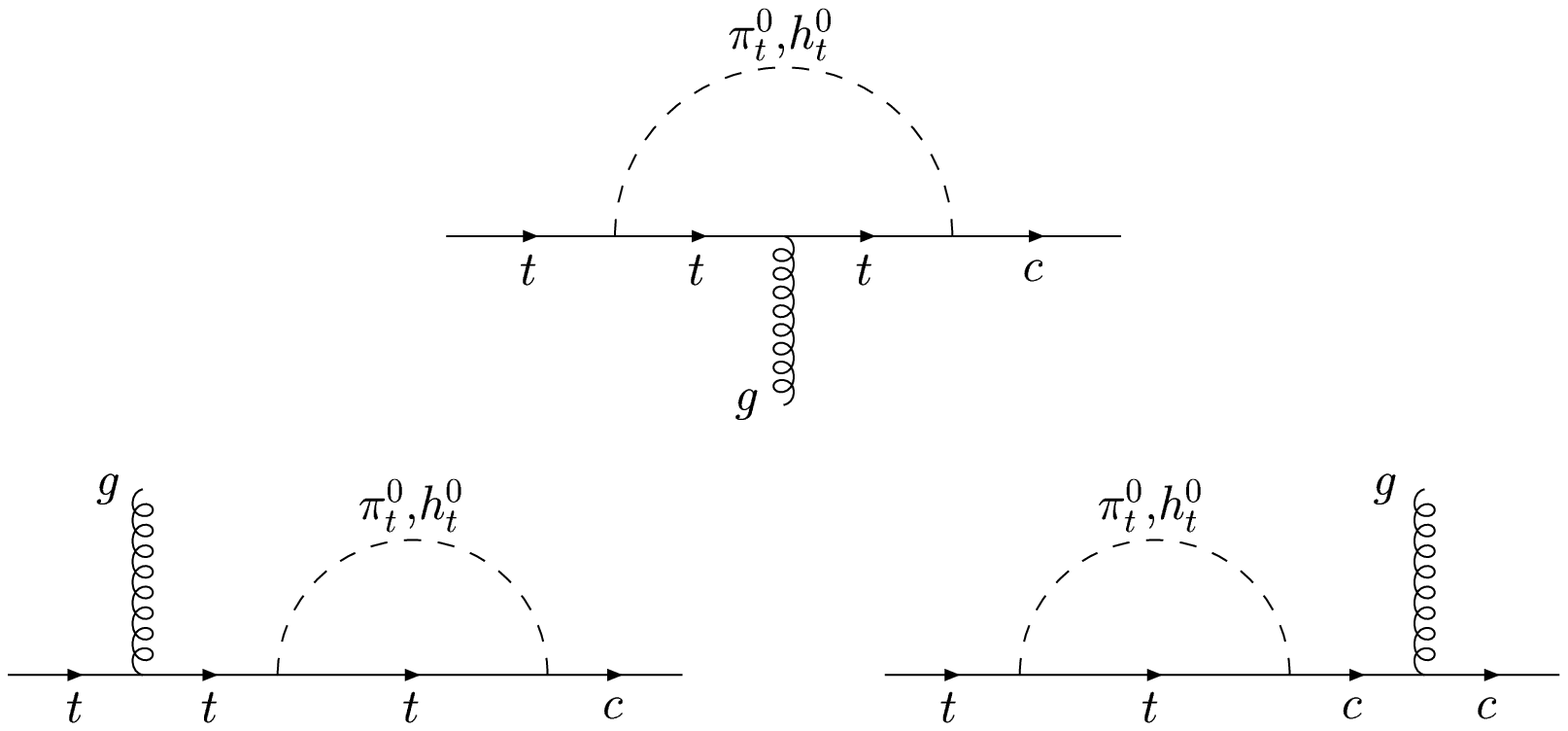}}
 \put(120,100){Fig.1}
\end{picture}
\end{center}
\end{figure}

\begin{figure}[hb]
\begin{center}
\begin{picture}(250,200)(0,0)
\put(-50,0){\epsfxsize120mm\epsfbox{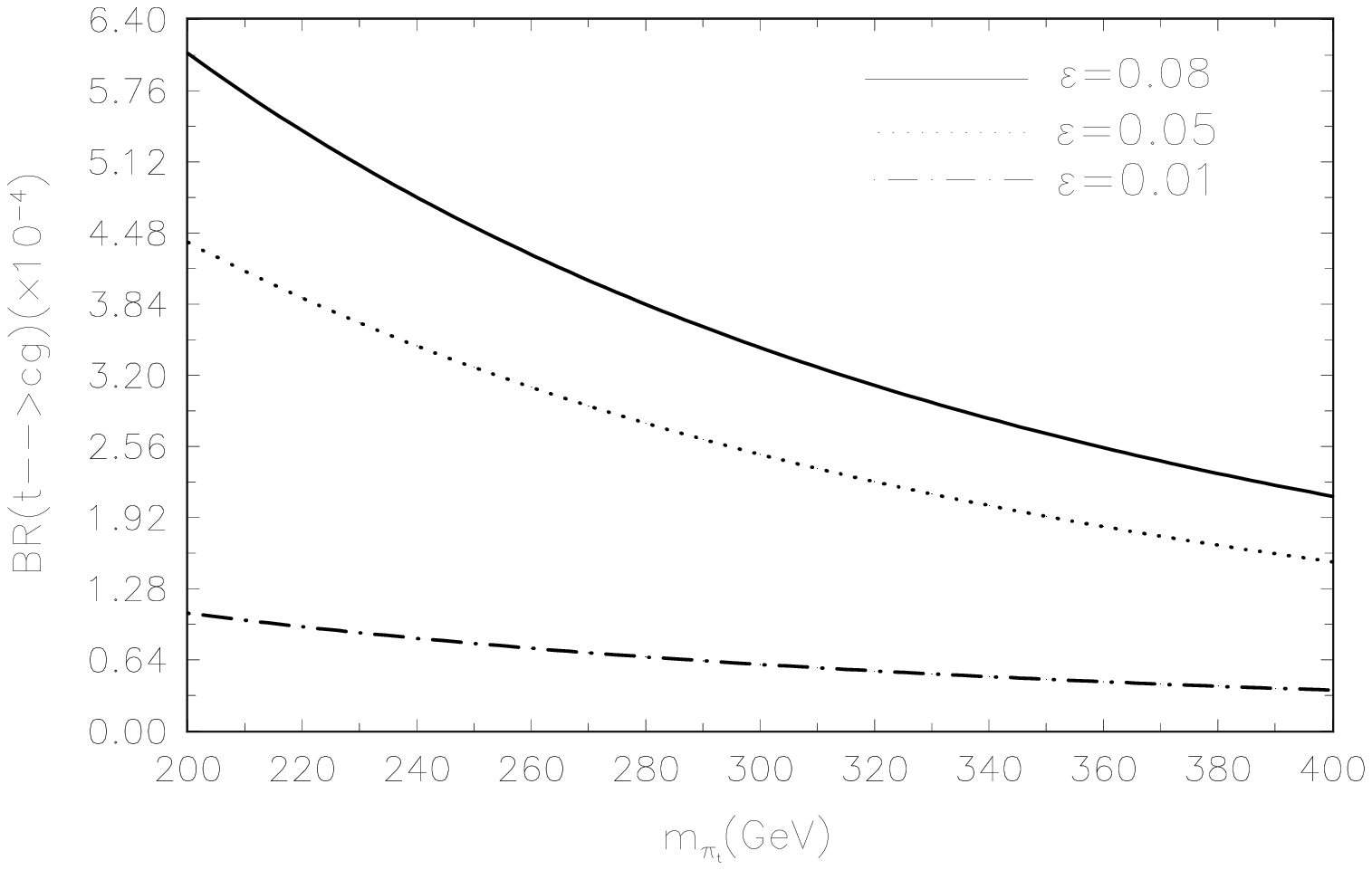}}
\put(120,-10){Fig.2}
\end{picture}
\end{center}
\end{figure}

\newpage
\begin{figure}[pt]
\begin{center}
\begin{picture}(250,200)(0,0)
\put(-50,0){\epsfxsize120mm\epsfbox{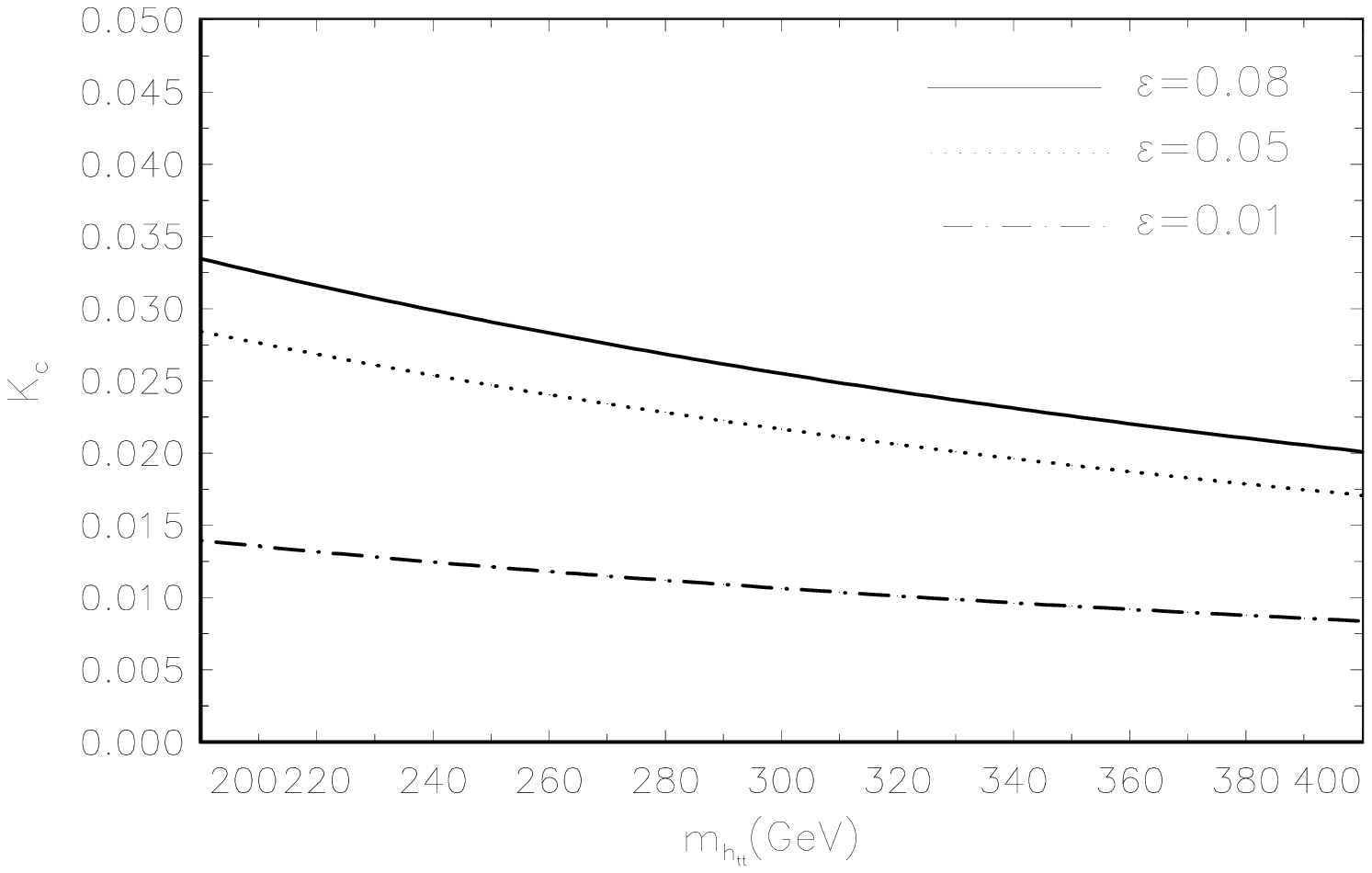}}
\put(120,-10){Fig.3}
\end{picture}
\end{center}
\end{figure}

\end{document}